\documentclass[12pt,preprint]{aastex}
\usepackage{emulateapj5}
\usepackage{apjfonts}
\usepackage{psfig}
\usepackage{amsmath}
\newcommand{\sfrac}[2]{\genfrac{}{}{}{1}{#1}{#2}}
\shorttitle{Transit Lightcurves}
\shortauthors{Mandel \& Agol}

\begin{document}


\title{Analytic Lightcurves for Planetary Transit Searches}
\shorttitle{Transit Lightcurves}

\author{Kaisey Mandel\altaffilmark{1,2} and Eric Agol\altaffilmark{1,3}}

\altaffiltext{1}{California Institute of Technology, Mail Code 130-33, 
Pasadena, CA 91125 USA}
\altaffiltext{2}{kmandel@tapir.caltech.edu}
\altaffiltext{3}{Chandra Fellow; agol@tapir.caltech.edu}

\slugcomment{}


\begin{abstract}
We present exact analytic formulae for the eclipse of a star described by 
quadratic or nonlinear limb darkening.  In the limit that the planet radius 
is less than a tenth of the stellar radius, we show that the exact lightcurve 
can be well approximated by assuming the region of the star blocked by the 
planet has constant surface brightness.  We apply these results to the HST 
observations of HD 209458, 
showing that the ratio of the planetary to stellar radii is $0.1207\pm 0.0003$.
These formulae give a fast and accurate means of computing lightcurves using 
limb-darkening coefficients from model atmospheres which should aid in the 
detection, simulation, and parameter fitting of planetary transits.
\end{abstract}



\section{Introduction}

The eclipse of the star HD 209458 by an orbiting planet was recently 
used to measure the size and mass of the planet, which had been 
found with velocity measurements \citep{cha00,hen00}.  
With this landmark discovery, the planetary transit tool was added to 
the planet-finder's toolbox,  already yielding several planetary candidates
\citep{uda02a,uda02b,dre02}.  Several large surveys which aim
to find planets using the transit signature are now being carried out or
planned and will soon yield large numbers of lightcurves requiring fast 
computation of eclipse models to find the transit needles within the 
haystack of variability \citep{bor01,how00, mal02,koc98,dee00,str02}.
Lightcurve fits to transit 
events may be used to characterize the planet and star, yielding 
important constraints on planet formation \citep{cod02,hub01,sea02}.  
The data require an accurate description of 
limb-darkening as demonstrated by Hubble Space Telescope 
observations of HD 209458 of such high quality that
a quadratic limb-darkening law was needed to fit the transit lightcurve 
rather than the usual linear limb-darkening law \citep{bro01}. 
The limb-darkening of 
main-sequence stars is represented by functions of 
$\mu=\cos{\theta}$, where 
$\theta$ is the angle between the normal to the stellar surface and 
the line of sight to the observer (Figure 1a).  \citet{cla00} has found 
that the most accurate limb-darkening functions are the quadratic law
in $\mu$ and the ``nonlinear" law which is a Taylor series to fourth 
order in $\mu^{1/2}$; the latter conserves flux to better than 0.05\%.

In this paper, we compute analytic functions for transit lightcurves
for the quadratic and nonlinear limb-darkening laws, and make available
our codes for the community ($\S$ \ref{conclusions}).
For treatment of subtler effects during planetary transits see 
\citet{sea00a,sea00b,hub01,hui02}.
In section 2 we review the lightcurve of a uniform spherical source.
In section 3 we derive the lightcurve for eclipses of nonlinear limb-darkened
stars.  In section 4 we give a simpler form in the limit of
a quadratic limb-darkening law.   In section 5 we give an approximation
for the lightcurve in the case $p\lesssim 0.1$ which is very fast to
compute and is fairly accurate.  In section 6 we apply the results to some 
example cases, and in section 7 we conclude.

\section{Uniform Source}

We model the transit as an eclipse of a spherical star by an opaque, dark
sphere.  In what follows, $d$ is the center-to-center distance between 
the star and the planet, $r_p$ is the radius of the planet, $r_*$ is 
the stellar radius, $z = d/r_*$ is the normalized separation of the 
centers, and $p = r_p/r_*$ is the size ratio (Figure 1b). The flux relative 
to the unobscured flux is $F$.  

For a uniform source, the ratio of obscured to unobscured flux is
$F^e(p,z) = 1-\lambda^e(p,z)$ where 
\begin{equation}\label{uniformoccult}
\lambda^e(p,z) = 
\left\{\begin{array}{ll}
0  &  1+p < z \\
{1 \over \pi} \left[p^2 \kappa_0+\kappa_1-\sqrt{{4z^2-(1+z^2-p^2)^2\over 4}}\right] &  |1-p| < z \le 1+p \\
 p^2 & z \le 1-p\\
1  &  z \le p-1,\\
\end{array}\right.
\end{equation}
and $\kappa_1=\cos^{-1}[(1-p^2+z^2)/2z],$ $\kappa_0=\cos^{-1}[(p^2+z^2-1)/2pz]$.
We next consider the effects of limb-darkening.

\section{Non-Linear Limb Darkening}

Limb darkening causes a star to be more centrally peaked in brightness compared
to a uniform source.  This leads to more significant dimming during 
eclipse and creates curvature in the trough.  Thus, including limb-darkening
is important for computing accurate eclipse lightcurves.
\citet{cla00} proposed a nonlinear limb-darkening law that fits well a wide
range of stellar models and observational bands,
$I(r) = 1 - \sum_{n=1}^4 c_n(1-\mu^{n/2})$,
where $\mu = \cos{\theta}=\sqrt{1-r^2}$, $0\le r \le 1$ is the 
normalized radial coordinate on the disk of the star and $I(r)$ is the 
specific intensity as a function of $r$ or $\mu$ with $I(0)=1$.  
Figure 1(a) shows the geometry of lensing and the definition of $\mu$.
The lightcurve in the limb-darkened case is given by 
\begin{equation}
F(p,z)= \left[\int_0^1 dr 2r I(r)\right]^{-1} \int_0^1 dr I(r) 
{d \left[F^e\left(p/r,z/r\right) r^2 \right]\over dr},
\end{equation}
where $F^e(p,z)$ is the lightcurve of a uniform source defined in the previous
section.

\centerline{\psfig{file=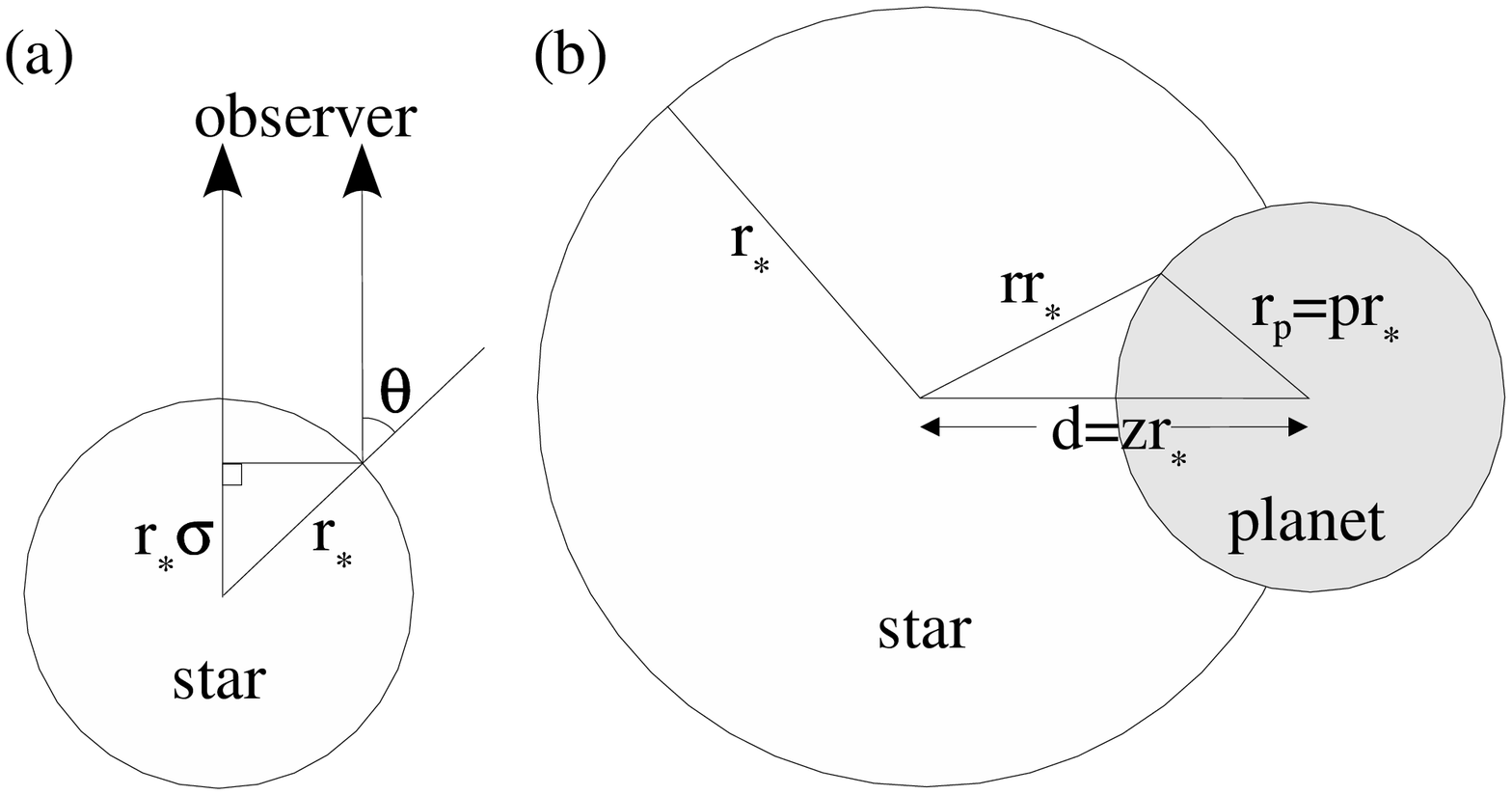,width=\hsize}} 
\figcaption{(a) Geometry of limb-darkening.  Star is seen edge on,
with the observer off the top of the page.  The star has radius
$r_*$ and $\theta$ is defined as the angle between the observer
and the normal to the stellar surface, while $\mu=\cos{\theta}$.
(b)  Transit geometry from perspective of observer.}

In what follows $c_0 \equiv 1-c_1 - c_2 - c_3 - c_4$.  
For convenience, we define 
$a \equiv (z-p)^2$, $b \equiv (z+p)^2$, and 
$\Omega = \sum_{n=0}^4 c_n(n+4)^{-1}$.  We partition the parameter space 
in $z$ and $p$ into the regions and cases listed in Table 1.
Next we describe each of these cases in turn.

In Case I the star is unobscured, so $F_I = 1$.  In Case II the planet disk 
lies on the limb of the star, but does not cover the center of the stellar 
disk.  We define
\begin{eqnarray}
N&=&\frac{(1-a)^{(n+6)/4}}{(b-a)^{1/2}}  B\left( \sfrac{n+8}{4}, 
\onehalf\right) \left[\frac{z^2 - p^2}{a} 
F_1\left(\onehalf,1,\onehalf,\sfrac{n+10}{4}; \sfrac{a-1}{a},
\sfrac{1-a}{b-a}\right) \right.\cr
 &-& \left. _2F_1\left(\onehalf,\onehalf;\sfrac{n+10}{4};
\sfrac{1-a}{b-a}\right)\right],
\end{eqnarray}
In the above equations, $B(a,b)$ is the Beta function, 
$F_1(a,b_1,b_2,c;x,y)$ is
Appell's hypergeometric function of two variables, and $ _2F_1(a,b,c;x)$ 
is the Gauss hypergeometric function.  The relative flux is 
$F = 1 - (2\pi\Omega)^{-1} \sum_{n=0}^4 N c_n(n+4)^{-1}$.
This case covers the ingress/egress where the lightcurve is steepest.  
For Cases III and IV,  the planet's disk lies entirely inside the stellar 
disk, but does not cover the stellar center. We define
\begin{eqnarray}
M&=& (1-a)^{(n+4)/4} \left[ \sfrac{z^2 - p^2}{a} 
F_1\left(\onehalf, -\sfrac{n+4}{4},1,1; \sfrac{b-a}{1-a},\sfrac{a-b}{a}\right)\right. \cr
&-&\left. _2F_1\left(-\sfrac{n+4}{4},\onehalf,1;\sfrac{b-a}{1-a} \right)
\right]
\end{eqnarray}
and $L= p^2\left(1 - p^2/2 - z^2\right)$.
Then the relative flux is given by
$F = 1 - (4\Omega)^{-1}\left[c_0 p^2 + 2\sum_{n=1}^3Mc_n(n+4)^{-1} + c_4 L\right]$.
This case requires the planet to be less than half of the size of the star.
In Case V the edge of the planet touches the center of the stellar disk, 
and the planet lies entirely within the stellar disk.  
The relative flux is $F_V = 1/2 + (2\Omega)^{-1} \sum_{n=0}^4 c_n(n+4)^{-1}
\,  _2F_1\left(1/2, -(n+4)/4,1;4p^2\right)$.
For Case VI the planet's diameter equals the star's radius, and the edge of the
planet's disk touches both the stellar center and the limb of the star.
The relative flux is
\begin{equation}
F = \frac{1}{2}+ \frac{1}{2\sqrt{\pi}\Omega} \sum_{n=0}^4 \frac{c_n}{n+4} 
\Gamma(\sfrac{3}{2} + \sfrac{n}{4})/\Gamma(2+\sfrac{n}{4}).
\end{equation}
In Case VII the edge of the planet's disk touches the stellar center, 
but the planet is not entirely contained inside the area of the stellar disk.
The relative flux is
\begin{equation}
F = \frac{1}{2}+ \frac{1}{4p\pi\Omega} \sum_{n=0}^4 \frac{c_n}{n+4} 
B\left(\onehalf, \sfrac{n+8}{4}\right) \,_2F_1\left(\onehalf,
\onehalf,\sfrac{5}{2} + \sfrac{n}{4}; \sfrac{1}{4p^2}\right).
\end{equation}
\centerline{\psfig{file=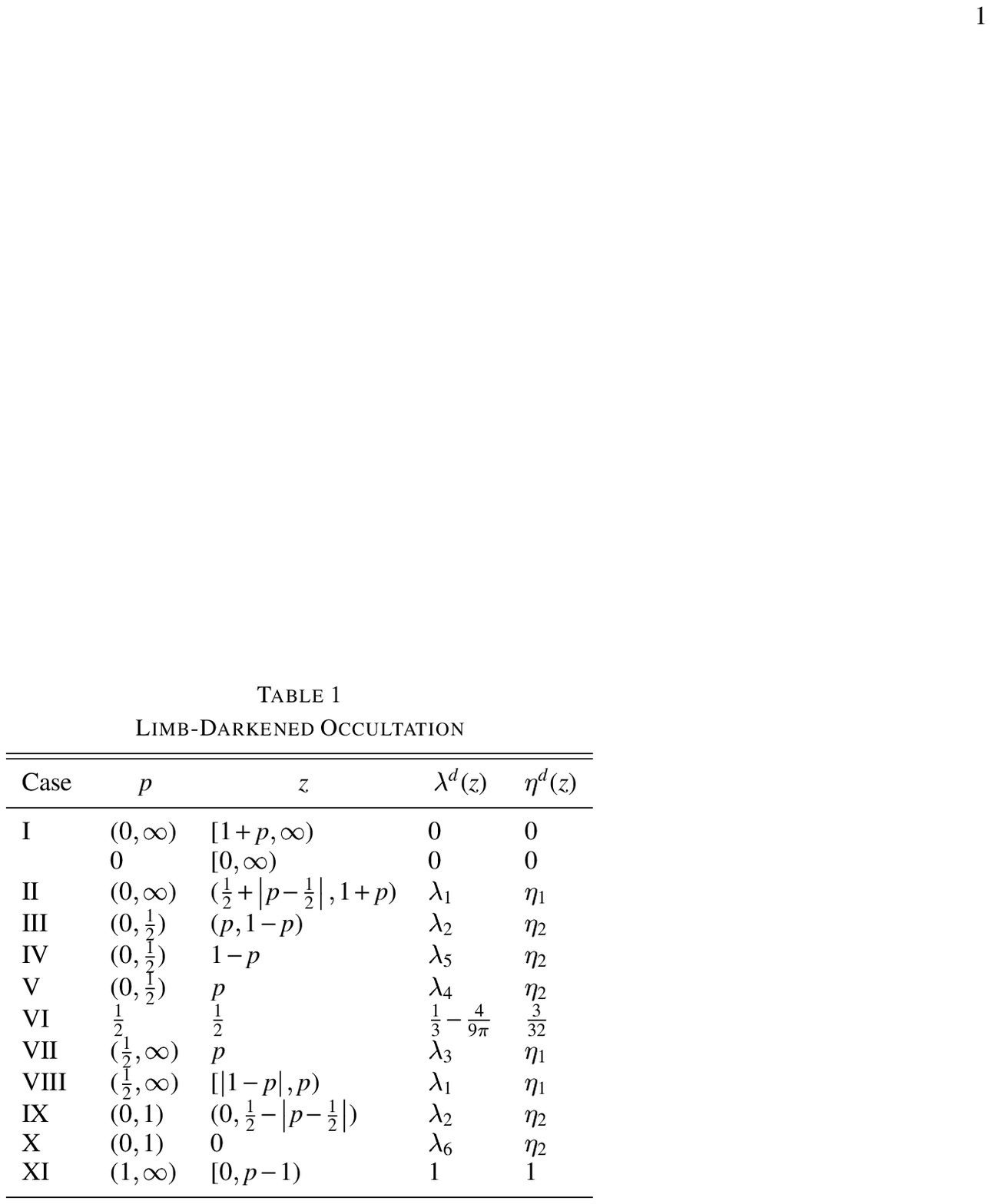,width=2.5in}} 

In Case VIII the planet covers the center and limb of the stellar disk.
The relative flux is $F = - \Omega^{-1} \sum_{n=0}^4 c_nN(n+4)^{-1}$.
This and the previous case apply when the planet is larger than 
half the size of the star.
For the special Case IX the planet's disk lies entirely inside the stellar disk, 
and the planet covers the stellar center.
The relative flux is
$F = (4\Omega)^{-1}[c_0(1-p^2)+c_4(1/2- L) 
- 2 \sum_{n=1}^3 c_n(n+4)^{-1} M]$.
This is the bottom of the transit trough for nearly edge-on
inclinations if $p \ll 1$.
In Case X the planet is concentric with the disk of 
the star, at the precise bottom of the transit trough.  In this case,
$F = \Omega^{-1} \sum_{n=0}^4 c_n (1-p^2)^{(n+4)/4}(n+4)^{-1}$.
This formula applies only for edge-on orbits when there is a central transit.
Finally, in Case XI the planet completely eclipses the star, so that 
$F = 0$.  In this case the ``planet'' is likely a star.
In the event that $c_1=c_3=0$, these lightcurves can be simplified
as we describe in the next section.

\section{Quadratic Limb Darkening}

In this section we describe the limb-darkening with a function which 
is quadratic in $\mu$, $I(r) = 1-\gamma_1(1-\mu)-\gamma_2(1-\mu)^2$,
where $\gamma_1+\gamma_2 < 1$.  The nonlinear law in the previous
section reduces to this case when $c_1=c_3=0$, $c_2=\gamma_1+2\gamma_2$, and 
$c_4=-\gamma_2$.  In this limit, the hypergeometric functions reduce to
elliptic integrals which are much faster to compute, so in this section
we provide these simpler formulae.

For a quadratic limb-darkening law, the lightcurve is 

$F = 1-\left(4\Omega\right)^{-1}\left[(1-c_2)\lambda^e+ c_2
\left(\lambda^d +\sfrac{2}{3}\Theta(p-z)\right)-c_4\eta^d \right]$,
where $\lambda^e$ is defined in Equation \ref{uniformoccult},
while $\lambda^d$ and $\eta^d$ are given in Table 1.

In Table 1, the various functions are
\begin{eqnarray}
\lambda_1&=& {1\over 9\pi\sqrt{pz}}\left[\vphantom{a-1\over a}\left((1-b)(2b+a-3)-3q(b-2)\right)K(k)  \right.\cr
 &+&\left.4pz(z^2+7p^2-4)E(k)-3{q\over a}\Pi\left(\sfrac{a-1}{a},k\right)\right],\cr
\lambda_2&=&{2\over 9 \pi\sqrt{1-a}}\left[\vphantom{a-b\over a}\left(1-5z^2+p^2+q^2\right)K(k^{-1}) \right.\cr
     &+&\left. (1-a)(z^2+7p^2-4)E(k^{-1})-3{q\over a}\Pi\left(\sfrac{a-b}{a},k^{-1}\right)\right],\cr
\lambda_3&=&{1\over 3}+{16p\over 9\pi}\left(2p^2-1\right)E\left(\sfrac{1}{2k}\right)
-{(1-4p^2)(3-8p^2) \over 9\pi p}K\left(\sfrac{1}{2k}\right),\cr
\lambda_4&=&{1\over 3}+{2\over 9\pi}\left[4\left(2p^2-1\right)E(2k)+\left(1-4p^2\right)K(2k)\right],\cr
\lambda_5&=&{2\over 3\pi}\cos^{-1}\left(1-2p\right)-{4\over 9\pi}(3+2p-8p^2),\cr
\lambda_6&=&-{2\over 3}\left(1-p^2\right)^{3/2},\cr
\eta_1 &=& (2\pi)^{-1}\left[\kappa_1+2\eta_2\kappa_0-\onequarter
\left(1+5p^2+z^2\right)
\sqrt{(1-a)(b-1)}\right],\cr
\eta_2 &=& {p^2 \over 2}\left(p^2+2z^2\right).
\end{eqnarray}
where $k=\sqrt{(1-a)/(4zp)}$ and $q=p^2-z^2$.
Here $\Pi(n,k)$ is the complete elliptic integral of the third kind with the
sign convention of \citet{gra94}.
For linear limb-darkening, $\gamma_2=0$, \citet{mer50} presents an
equivalent analytic expression in terms of an ``eclipse function,'' $\alpha$.
The expressions here require fewer evaluations of the elliptic integrals,
which decreases computation time, and include quadratic limb darkening.  
Our expression for eclipse with quadratic limb-darkening decreases computation 
time by more than an order of magnitude compared to evaluating the expressions 
in the previous section or numerical integration of the unocculted flux.  

\section{Small Planets}

For a small planet, $p \lesssim 0.1$, the interior of the lightcurve, $z < 1-p$, 
can be approximated by assuming the surface-brightness of the star is
constant under the disk of the planet, so that
$F = 1-{p^2 I^*(z)\over 4\Omega}$, and
$I^*(z)=(4zp)^{-1}\int_{z-p}^{z+p} I(r)2r dr$.
If one knows the limb-darkening coefficients of
the star in question (from, say, spectral information), and if
the semi-major axis is much larger than the size of the star
so that the orbit can be approximated by a straight
line, then the shape of the eclipse for $p \lesssim 0.1$ is simply determined 
by the smallest impact parameter, $z_0=a_p\cos{i}/r_*$, where
$a_p$ is the semi-major axis and $i$ is the inclination angle. For example,
at the midpoint of the eclipse, $z=z_0$, while at the 1/4 and 3/4 
phases of the eclipse, $z=z_{1/4}=(1+3z_0^2)^{1/2}/2$.
Taking the ratio of the depth of the eclipse at these points
yields $R=(1-F(z_0))/(1-F(z_{1/4}))=I(z_0)/I(z_{1/4})$.
This then determines an equation for $z_0$, and the resulting
$z_0$ can then be used to determine $p=\sqrt{4\Omega(1-F(z_0))/I(z_0)}$.
When $1-p < z < 1+p$ and $p \lesssim 0.1$ an approximation to the lightcurve is 
\begin{eqnarray} \label{approx2}
F &=& 1-{I^*(z)\over 4\Omega} \left[p^2 \cos^{-1}\left({z-1\over p}\right)
-(z-1)\sqrt{p^2-(z-1)^2}\right],
\end{eqnarray}
where $I^*(z)=(1-a)^{-1}\int_{z-p}^1 I(r)2r dr$.
which is accurate to better than 2\% of $1-F(0)$ for $p=0.1$ and 
$\Sigma_{n=1}^4 c_n \le 1$ (see Figure 2).
This is a very fast means of computing transit lightcurves with
reasonable accuracy, and may be used for any limb-darkening function,
generalizing the approach of \citet{dee01}.

\section{Discussion}

Figure 2 shows five lightcurves, the first of which has 
$c_n=0, \{n=1,4\}$, while the other four have $c_n=1, c_m=0,
\{m\ne n\}$ for $p=0.1$;  these may be thought of as a
basis set for any nonlinear limb-darkening.
Note that the higher order functions have flux which is
concentrated more strongly toward the center of the star, and thus
have a more gradual ingress and a deeper minimum as more flux is blocked 
at the center than the edge.  All of the curves cross near $z \sim 0.7$,
which means that accurate observations are required near minimum
and egress/ingress to constrain the coefficients of the various
basis functions.  If the inclination is large enough that $z \gtrsim
0.7$ for the entire transit, then it may be difficult to constrain
the $c_n$'s.
 
\centerline{\psfig{file=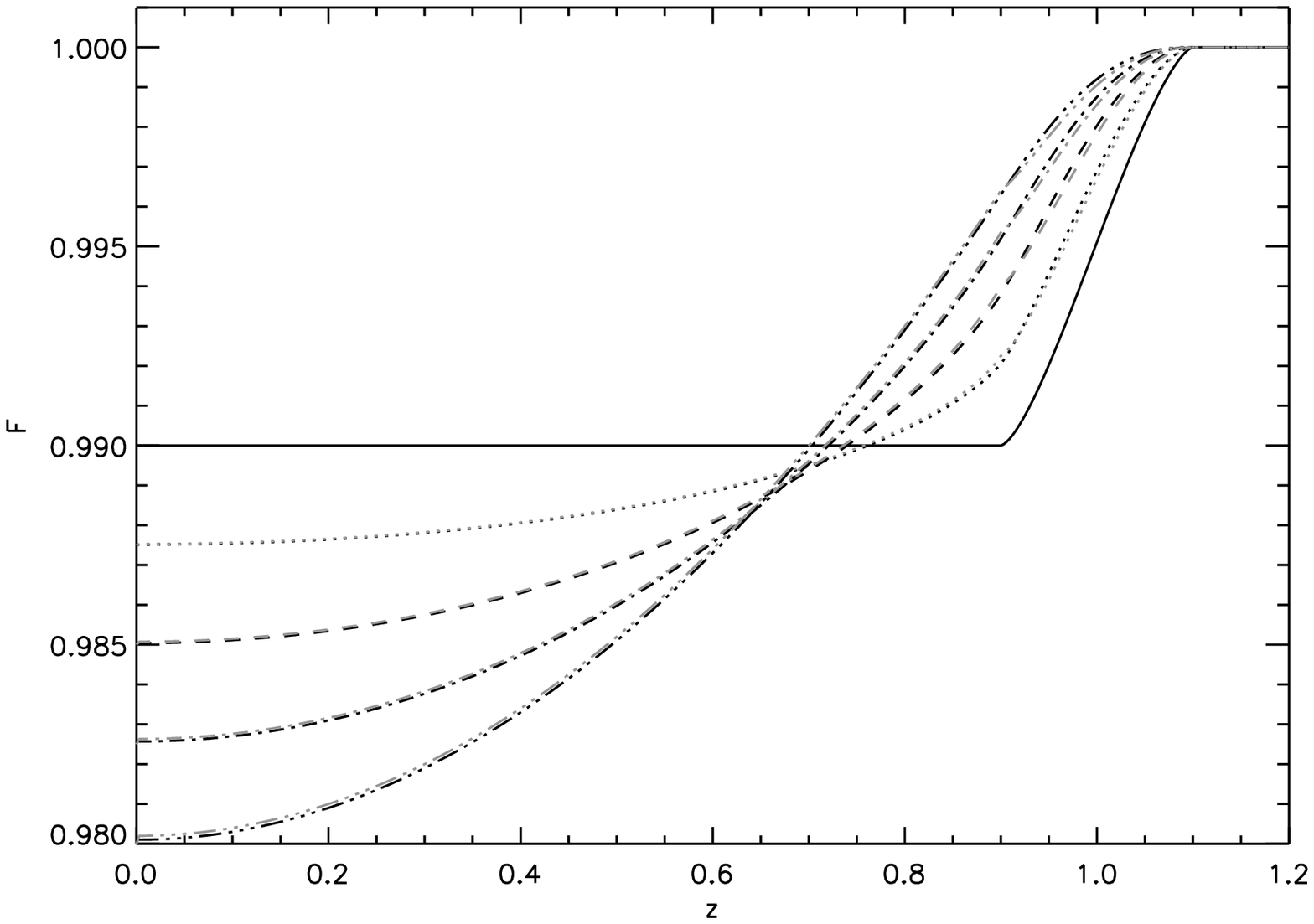,width=\hsize}} 
\figcaption{Transit lightcurves for $p=0.1$ and $c_1=c_2=c_3=c_4=0$
(solid line), and all coefficients equal zero but $c_1=1$ (dotted line), 
$c_2=1$ (dashed line), $c_3=1$ (dash-dot line), or $c_4=1$ (dash, 
triple-dot line). The lighter lines (nearly indistinguishable) 
show the approximation of section 5.}

As \citet{cla00} claims that the nonlinear limb-darkening law is
the most accurate, we have compared the transit with
$p=0.1$ for the nonlinear and quadratic laws.  
Of the entire grid of models computed by \citet{cla00}
which covers 2000 K $<$T$_{eff} <$ 50000 K, 0 $<\log{\rm g}<5$, 
-5 $<$ [M/H]$<1$, and filters u, b, v, y, U, B, V, R, I, J, H, K, 
the largest difference between the quadratic and nonlinear models 
for $p=0.1$ is 3\% of the maximum of $1-F$.  
Thus, the quadratic law should be 
sufficient for main-sequence stars when an accuracy of less than
3\% is required; indeed, the average difference for the entire
grid of models is about 1\% of the maximum value of $1-F$ for
each lightcurve.  In absolute terms, this is about $10^{-4}(p/0.1)^2$
of the total flux, an accuracy which can be achieved from space.

So far we have only presented the lightcurve as a function of $z$ and $p$.
To determine $z$ as a function of time requires the planetary orbital
parameters, which for zero eccentricity is given as
$z = a_pr_*^{-1}\left[\left(\sin{\omega t}\right)^2+
\left(\cos{i}\cos{\omega t}\right)^2\right]^{1/2}$,
where $\omega$ is the orbital frequency, while $t$ is the time
measured from the center of the transit.  Contribution of flux
from the planetary companion or other companions may be added
to the lightcurve, reducing the transit depth.

To illustrate the utility of our formulae, we have fit the 
nonlinear limb-darkened lightcurve to the HST-STIS data of 
HD 209458 \citep{bro01}.  The best-fit parameters in
this case are $p=0.12070\pm 0.00027$, $i=86.591^\circ\pm 0.055$,
$a_p/r_*=8.779\pm 0.032$, $c_1=0.701$, $c_2=0.149$, $c_3=0.277$,
$c_4=-0.297$, with a reduced $\chi^2=1.046$.  The $c_n$'s are 
poorly constrained given the small differences in the basis
functions relative to the observed errors and the large impact 
parameter for this system.  The errors on the other parameters 
are marginalized over the $c_n$'s.  Limiting
the limb-darkening to quadratic, we find $\gamma_1=0.296\pm 0.025$
and $\gamma_2=0.34\pm0.04$, consistent with the values derived
by \citet{bro01} and with stellar atmosphere predictions.
The value for $p$ in the quadratic case is consistent with
the nonlinear case, indicating that the fit is independent of
the assumed limb-darkening law.  For a stellar mass of 
$1.1\pm0.1 M_\odot$ \citep{maz00} and period of $T=3.5248$ days 
\citep{bro01}, we find $r_*=1.145\pm 0.035 R_\odot$ and $r_p=1.376\pm
0.043 R_{Jup}$.

We apply the small-planet approximation described in the previous
section to HD 209458, assuming quadratic limb-darkening.  \citet{cod02}
determined the effective temperature and surface gravity for this
star,  which imply $\gamma_1=0.292$ and $\gamma_2=0.35$ for the 
I-band flux (close to the effective wavelength of the HST STIS 
data) from the models of \citet{cla00}. From
the lightcurve, one finds $F(z_0)=0.9835$ and $F(z_{1/4})=0.9847$.  
Solving the equations from Section 5 gives $z_0=0.546$ and $p=0.12$, 
very similar to the parameters derived by a fit to the entire lightcurve.  
This technique may be used for finding initial parameters for 
lightcurve fitting.

\section{Conclusions} \label{conclusions}

We have derived analytic expressions for an eclipse including quadratic 
limb-darkening and nonlinear limb-darkening.  The nonlinear law
(Section 3) provides an accurate fit to realistic stellar limb-darkening, while
the quadratic fit (Section 4) provides a fast means of obtaining a relatively
accurate lightcurve.  For an extremely fast and fairly accurate
approximation for any limb-darkening law, the equations in Section 5
may be used to derive lightcurves.  If the limb-darkening law is
known from the spectral type of the star, then one can use the
formulae in Section 5 to analytically estimate both
the minimum impact parameter (in units of stellar radius) and
the ratio of the planetary radius to the stellar radius.
We have written a code which takes the properties of a host star, finds
the limb-darkening coefficients in the tables of \citet{cla00}, and
computes lightcurves for the parameters of a given planetary transit.
This code will be useful for simulating planetary transit searches
\citep{gau00,def01,jen02a,rem02,jen02b,pep02},
searching for planetary transit signals in lightcurves collected by
a given search, and for fitting and measuring the errors of the
parameters of detected planetary transit events.   Planetary searches 
suffer from two important backgrounds:  grazing eclipsing binaries and 
triple systems in which two stars eclipse while the flux from the 
third reduces the depth of the eclipse.  Using the appropriate 
limb-darkening coefficients for each star's spectral type will help 
to distinguish these contaminants from true planetary transits, which 
can be accomplished using the formulae presented here.
These routines are made available by download
from http://www.pha.jhu.edu/$\sim$agol/.

\acknowledgments

We thank Sara Seager and Leon Koopmans for useful discussions.
Support for E.A. was provided by the National Aeronautics and Space Administration 
through Chandra Postdoctoral Fellowship Award PF0-10013 issued by the Chandra 
X-ray Observatory Center, which is operated by the Smithsonian Astrophysical 
Observatory for and on behalf of the National Aeronautics Space Administration 
under contract NAS 8-39073.  K.M. was supported by a Caltech Summer Undergraduate
Research Fellowship.

\end{document}